\documentclass[aps,prl,twocolumn,groupedaddress,amsmath,amssymb]{revtex4}
\usepackage{amssymb}
\usepackage{amsmath}
\usepackage{graphicx}
\usepackage{subfigure}
\usepackage{textcomp}
\usepackage{color}
\usepackage{amsfonts}
\usepackage{bbold}
\usepackage{dsfont}
\usepackage{epsfig}
\usepackage{hyperref}

\begin{document}

\title{A pertubative approach to the Kondo effect in magnetic atoms on nonmagnetic substrates}

\author{Aaron Hurley, Nadjib Baadji and Stefano Sanvito}
\affiliation{School of Physics and CRANN, Trinity College, Dublin 2, Ireland}

\date{\today}

\begin{abstract}
Recent experimental advances in scanning tunneling microscopy make the measurement of the conductance spectra of isolated 
and magnetically coupled atoms on nonmagnetic substrates possible. Notably these spectra are characterized by a competition 
between the Kondo effect and spin-flip inelastic electron tunneling. In particular they include Kondo resonances and a logarithmic enhancement 
of the conductance at voltages corresponding to magnetic excitations, two features that cannot be captured by second order perturbation 
theory in the electron-spin coupling. We have now derived a third order analytic expression for the electron-spin self-energy, which can be 
readily used in combination with the non-equilibrium Green's function scheme for electron transport at finite bias. We demonstrate that our 
method is capable of quantitative description the competition between Kondo resonances and spin-flip inelastic electron tunneling
at a computational cost significantly lower than that of other approaches. The examples of Co and Fe on CuN are discussed in detail.
\end{abstract}


\maketitle

The interaction between conduction electrons and localized spins in transition metals with partially filled $d$ shells is central to many 
low-temperature spin effects, which may underpin the development of spintronics and quantum information devices. When adsorbed on the 
surface of a metallic host, magnetic transition metal atoms exhibit various distinctive features in the conductance spectrum, which are indicative 
of many-body scattering between the conduction electrons and the localised spins. These manifest themselves as conductance steps at voltages 
corresponding to the quasi-particle energies of specific magnetic excitations and as zero-bias conductance peaks, known as Kondo resonances.
The first are associated to spin-flip inelastic electron tunneling and can be described by second order perturbation theory in the electron-spin 
coupling~\cite{Hurley}, but the second results from third order effects due to the electron screening of the local spins.

Recent advances in scanning tunneling microscopy (STM) have enabled the detection of many-body scattering events in
Mn~\cite{Hir1}, Fe~\cite{Hir2} and Co~\cite{Otte1,Otte2} adatoms adsorbed on a CuN insulating substrate. The reduced symmetry of the surface 
leads to significant magnetic anisotropy, especially for Fe and Co. Fe is found to have a large easy-axis anisotropy [$D<0$, see Eq.~(\ref{eq:2})], 
leading to a ground state spin close to that of the maximum $z$-component of the integer $S=2$. This results in four evenly spaced conductance 
steps in the spectrum. For Co the large hard-axis anisotropy ($D>0$) and the half-integer $S=3/2$ spin produce a doublet ground state.  
The measured zero-bias Kondo resonance is then due to spin transitions between the degenerate ground state levels.

Theoretical attempts to reproduce these conductance spectra have focused largely on including second order scattering events, which 
cannot account for Kondo resonances but fare well in reproducing the conductance steps and their relative intensities 
\cite{Fernandez-Rossier,Fransson,Lorente,Delgado,Sothmann,Hurley,Persson}. Addressing Kondo physics in Co is more involved and 
one has to look for alternative techniques, such as density functional theory (DFT) informed numerical renormalisation group 
\cite{Zitko1,Zitko2}. These schemes however are numerically expensive. In this letter we extend the perturbative approach to the third 
order in the electron-spin scattering and derive an analytic expression for the scattering self-energy. This is then implemented within the 
non-equilibrium Green's function (NEGF) formalism \cite{Datta,Rocha} for electron transport and used to calculate the STM conductance 
spectra of both Fe and Co on CuN.

The Hamiltonian describing the STM tip, the magnetic adatom and the nonmagnetic substrate can be divided into three components, 
namely a purely electronic part,  $H_\mathrm{e}$, a purely spin part, $H_\mathrm{sp}$, and an electron-spin interaction part, $H_\mathrm{e-sp}$.
These are given respectively by
\begin{align}
\label{eq:1}
&{H}_\mathrm{e}=\sum_{kl}\varepsilon_{kl}a_{kl}^{\dagger}a_{kl}+\varepsilon_0\sum_{\alpha}c_{\alpha}^{\dagger}c_{\alpha}+H_\mathrm{tun}\:,\\
\label{eq:2}
&{H}_\mathrm{sp}=D({S}^z)^2+E[({S}^x)^2-({S}^y)^2]\:,\\
\label{eq:3}
&{H}_\mathrm{e-sp}=J_\mathrm{sd}\sum_{\alpha,\alpha'}(c_{\alpha}^{\dagger}[\boldsymbol{\sigma}]_{{\alpha}{\alpha'}}c_{\alpha'})\cdot\mathbf{S}\:.
\end{align}
In $H_\mathrm{e}$ we assume that both the tip (t) and the substrate (s) are characterized by a single $s$-like band, so that the electronic
structure (spin-degenerate) is described by the creation (annihilation) operators $a_{kl}^{\dagger}$ ($a_{kl}$) and the energies $\varepsilon_{kl}$ 
($l=\{\mathrm{t},\mathrm{s}\}$). Similarly the operators $c_{\alpha}^{\dagger}$/$c_{\alpha}$ and the onsite energy 
$\varepsilon_0$ define the electronic level of the adatom. The tunneling between the adatom and the tip/substrate is accounted for by $H_\mathrm{tun}$. 
This term can be replaced by an effective energy-dependent self-energy $\Sigma_\mathrm{t,s}(E)$, which produces a constant broadening of the 
adatom density of states given by $\Gamma_\mathrm{t,s}=\gamma_\mathrm{t-a/s-a}^2/W$ \cite{Hurley}. The broadening is thus constant around 
the Fermi energy ($E_\mathrm{F}$) provided that the bandwidth, $W$, is large compared to the coupling strength between the two leads and the 
adatom (a), $\gamma_\mathrm{t-a/s-a}$. $H_\mathrm{sp}$ contains information on the magnetic anisotropy of the adatom and includes both
an axial ($D$) and a transverse ($E$) term. $H_\mathrm{e-sp}$ couples the conducting electrons to the local spin $\mathbf{S}$ through an 
exchange interaction parameter $J_\mathrm{sd}$.

The calculation of the spin-inelastic conductance spectra of adatoms adsorbed on surfaces requires the implementation 
of the NEGF scheme including many-body scattering effects, which has been rigorously derived in reference \cite{Hurley}. 
In brief, the tunneling current at any terminal $i=\mathrm{t,s}$ and bias voltage $V$ can be calculated from
\begin{align}
\label{eq:29}
&I_i=\int_{-\infty}^{+\infty}\bar{I_{i}}(E)\:dE\:,\\
\label{eq:30}
\bar{I_{i}}(E)=\frac{e}{h}\text{Tr}\{&[\Sigma_{i}^{>}(E)G^{<}(E)]-[\Sigma_{i}^{<}(E)G^{>}(E)]\},
\end{align}
where $\Sigma_{i}^{>}(E)=[1-f_i(E,V)]\Gamma_i$ and $\Sigma_{i}^{<}(E)=f_i(E,V)\Gamma_i$. Here $f_i(E,V)$ is the bias dependent 
Fermi function in each of the leads, $e$ is the electron charge and $h$ the Planck constant. The many body lesser/greater Green's 
functions $G^{\lessgtr}(E)$ require the calculation of a lesser/greater interacting self-energy $\Sigma^{\lessgtr}_\mathrm{int}(E)$ which 
is derived next.

Let us consider a nonequilibrium system at finite temperature described at the level of the Keldysh \cite{Keldysh,Haug} formalism. 
The contour-ordered single particle many-body Green's function is defined as
\begin{equation}
\label{eq:4} [G(\tau,\tau')]_{\sigma\sigma'}=-i{\langle}|T_C\{c_{\sigma}(\tau)c^{\dagger}_{\sigma'}(\tau')\}|{\rangle}.
\end{equation}
This can expanded up to the $n$-th order \cite{Mahan} in the electron-spin coupling 
\begin{align}
\label{eq:5}
&[G(\tau,\tau')]_{\sigma\sigma'}=\sum_n\frac{(-i)^{n+1}}{n!}\int\limits_C{d}\tau_1\dots\int\limits_C{d}\tau_n\ \times \nonumber \\ &\frac{{\langle}0|T_C
\{{H}_\mathrm{e-sp}(\tau_1)\dots{H}_\mathrm{e-sp}(\tau_n)c_{\sigma}(\tau)c_{\sigma'}^{\dagger}(\tau')\}|0{\rangle}}{U(-\infty,-\infty)},
\end{align}
where $U$ is the time-evolution unitary operator and the time-averages are now performed over the non-interacting ground state $|0{\rangle}$. 

Exact diagonalization is performed for $H_\mathrm{sp}$ [Eq.~(\ref{eq:2})] yielding the spin energies $\varepsilon_m$ and the states
$|m\rangle$ ($\Omega_{mn}=\varepsilon_m-\varepsilon_n$). Then by a careful evaluation of the Feynman integrals in Eq.~(\ref{eq:5}) 
we arrive at an analytic expression for the interacting greater and lesser self-energies, $\Sigma^{\lessgtr}_\mathrm{int}$, which includes both 
contributions to the second and the third order in $H_\mathrm{e-sp}$ (note that diagrams containing fermion loops vanish due to the electron-spin 
selection rules \cite{Mahan}). This reads ($+$ corresponds to $<$ and $-$ to $>$)
\begin{widetext}
\begin{align}
\label{eq:6}
&\Sigma^{\lessgtr}_\mathrm{int}(E)=J^2_\mathrm{sd}\sum_{m,n,l}P_l(1-P_m)G^{\lessgtr}_0(E\pm\Omega_{ml})\Big\{\delta_{nl}\sum_{i}|{\langle}m|S^i|n{\rangle}|^2 \nonumber \\
&-i({\rho}J_\mathrm{sd})\sum_{ijk}\varepsilon_{ijk}{\langle}m|S^i|n{\rangle}{\langle}n|S^j|l{\rangle}{\langle}l|S^k|m{\rangle}
\Big[\text{ln}\Big|\frac{W}{\sqrt{(E\pm\Omega_{mn})^2+(k_\mathrm{B}T)^{2}}}\Big|+\text{ln}\Big|\frac{W}{\sqrt{(E\pm\Omega_{nl})^2+(k_\mathrm{B}T)^{2}}}\Big|\Big]\Big\}\:,
\end{align}
\end{widetext}
where $k_\mathrm{B}$ is the Boltzmann constant, $T$ is the temperature and $P_l$ is the population of the $|l\rangle$ spin state. 
If we now assume that the adatom is much more strongly coupled to the substrate than to the STM tip $(\gamma_\mathrm{s}\gg\gamma_\mathrm{t})$, 
we can approximate its density of states around $E_\mathrm{F}$ with a constant, $\rho=(\Gamma_s/2\pi)/(\varepsilon_0^2+\Gamma_s^2/4)$, where 
$\varepsilon_0\gg E_\mathrm{F}$. The weak coupling to the STM tip also ensures that the spin system remains always close to equilibrium \cite{Sothmann}. 
This means that the adatom state always resides close to the ground state, i.e. that $P_{0}{\sim}1$. The matrix elements ${\langle}m|S^i|n{\rangle}$, 
with $i=\{x,y,z\}$, determine the intensity of a given transition between an initial state $n$ and a final state $m$. We also note that the strength of the 3rd 
order interaction depends on the dimensionless parameter ${\rho}J_\mathrm{sd}$ in agreement with the expression of Ref.~\cite{Elste}.


Having found a close formula for the interacting self-energy we can now proceed with the full NEGF formalism \cite{Hurley} 
and calculate the conductance spectra of various adatoms on insulating substrates. In order to explore the Kondo physics 
contained in the 3rd order contributions to $\Sigma^{\lessgtr}_\mathrm{int}$, we consider both Co and Fe 
adsorbed on CuN, for which we can compare with experiments \cite{Hir2,Otte1,Otte2}. 

\begin{figure}[t]
\centering
\resizebox{\columnwidth}{!}{\includegraphics[width=5cm,angle=-90]{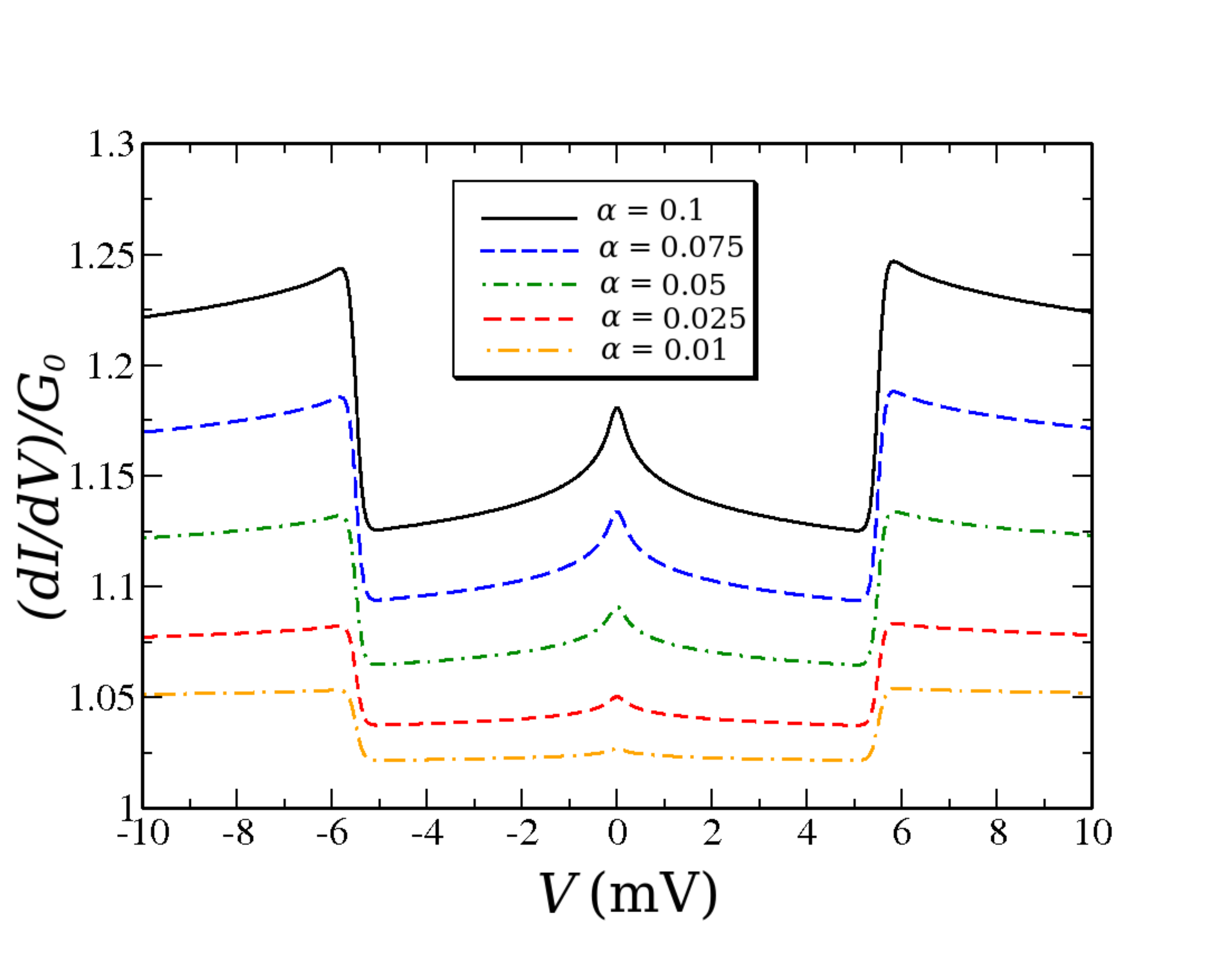}}
\caption{\footnotesize{Normalised conductance spectrum for Co ($S=3/2$) on CuN as $\alpha={\rho}J_\mathrm{sd}$ is increased. 
Note the emergence of a Kondo resonance at zero-bias, i.e. at the Fermi level.}}
\label{1}
\end{figure}

\begin{figure}[t]
\centering
\resizebox{\columnwidth}{!}{\includegraphics[width=5cm,angle=-90]{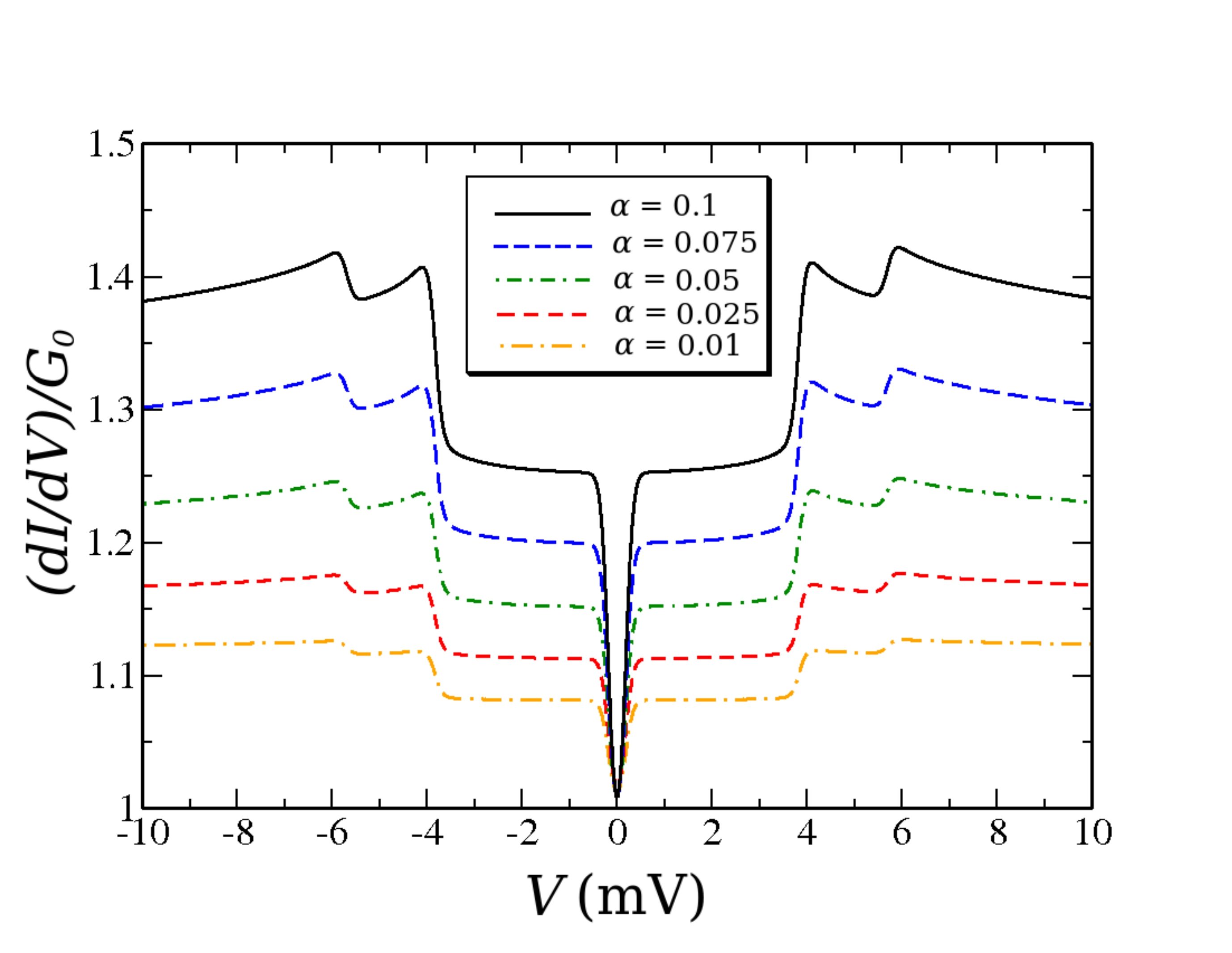}}
\caption{\footnotesize{Normalised conductance spectrum for Fe ($S=2$) on CuN as $\alpha={\rho}J_\mathrm{sd}$ is increased. Note the transition from 
a constant conductance at the inelastic steps to a conductance logarithmic decaying.}}
\label{2}
\end{figure}

Figures~\ref{1} and \ref{2} show the calculated conductance spectra respectively for single Co and Fe atoms on CuN as the dimensionless 
parameter $\alpha={\rho}J_\mathrm{sd}$ is varied (note that the spectra are normalized relatively to the elastic only conductance, $G_0$). 
Many of the parameters of the model used in the present simulations have been extracted from the DFT calculations \cite{Zitko2}. 
The broadening due to the coupling to the substrate (we neglect that due to STM tip) is set at $\Gamma_\mathrm{s}=0.1\:\text{eV}$ for both Co 
and Fe. In order to ensure consistency in our approximations ($\Gamma_\mathrm{s}=\gamma_\mathrm{s-a}^2/W$) we also set 
$\gamma_\mathrm{s-a}=1.5\:\text{eV}$ and the substrate bandwidth to $W=20\:\text{eV}$. We then choose $\varepsilon_0=1\:\text{eV}$ to 
fulfill the criterion $\varepsilon_0\gg E_\mathrm{F}=0$. The magnitude of $J_\mathrm{sd}$ for both Fe and Co is held constant at 0.5~eV 
(note that $J_\mathrm{sd}$=0.5~eV is comparable to what calculated from DFT for a Ni impurity sandwiched between Au leads \cite{Lucignano}). 
Finally the empirical parameters describing the axial, $D$, and translational, $E$ anisotropies are taken from Refs.~\cite{Hir2,Otte1} and 
are $D_{\text{Co}}=2.75~\text{meV}$, $E_{\text{Co}}=0~\text{meV}$, $D_{\text{Fe}}=-1.53~\text{meV}$ and $E_{\text{Fe}}=0.31~\text{meV}$, 
while the adsorbed atoms spins are $S_\text{Co}=3/2$ and $S_\text{Fe}=2$. 

The parameter $\alpha$ is varied by altering the coupling of the adatom to the substrate, $\gamma_\mathrm{s-a}$, thus by changing the value of the 
density of states at the Fermi level, $\rho$.
A full diagonalisation of $H_\mathrm{sp}$ for Co gives a set of four ($2S_\text{Co}+1$) eigenvalues and eigenvectors. In particular the presence 
of hard-axis anisotropy results in the following energy manifold $\varepsilon^\text{Co}_m=\{0.69, 0.69, 6.19, 6.19\}~\text{meV}$, i.e. in a doubly
degenerate ground state. It is then found that transitions between the degenerate ground states only become allowed on inclusion of the third 
order term in Eq.~(\ref{eq:6}) because of the selection rules imposed by the theory through the matrix elements ${\langle}m|S^i|n{\rangle}$. 
Such a transition appears in the spectrum of figure~\ref{1} in the form of a zero-bias Kondo peak as the value of $\alpha$ is increased. 
The greater is the density of states of the adatom, the greater becomes the third order Kondo contribution to the conductance. This also produces
a logarithmic rise in conductance both for the ground and for the first excited state transitions (detected at around 6~mV). These features 
matche closely what is observed in the experiments of Refs.~\cite{Otte1,Otte2}. A notable difference is that our calculation seem to show a 
peak considerably smaller than that in experiments. Such a discrepancy may arise from two-body electronic (Hubbard-like) contributions to 
the Hamiltonian, which we neglect here.

Results for Fe are presented in Fig. \ref{2}. This time the five eigenvalues of $H_\mathrm{sp}$ are 
$\varepsilon^\text{Fe}_m=\{-6.30, -6.12, -2.46,-0.60, 0.18\}~\text{meV}$, so that the ground state is non-degenerate. At zero magnetic 
field all transitions available by the 3rd order expansion are resolved. As $\alpha$ is increased the 3rd order Kondo term produces a logarithmic 
peak at each of the allowed conductance steps. This feature agrees well with experiments and it was previously explained by invoking a 
non-equilibrium population of the Fe spin states \cite{Sothmann}. This explanation however conflicts with the fact that the same effect is seen 
also for low currents \cite{Hir2}, namely when the adatom spin state resides close to the equilibrium. Therefore, we re-interpret the experimental 
data as the manifestation of the 3rd order Kondo effects in the spin-flip inelastic tunneling spectra.

\begin{figure}[t]
\centering
\resizebox{\columnwidth}{!}{\includegraphics[width=5cm,angle=-90]{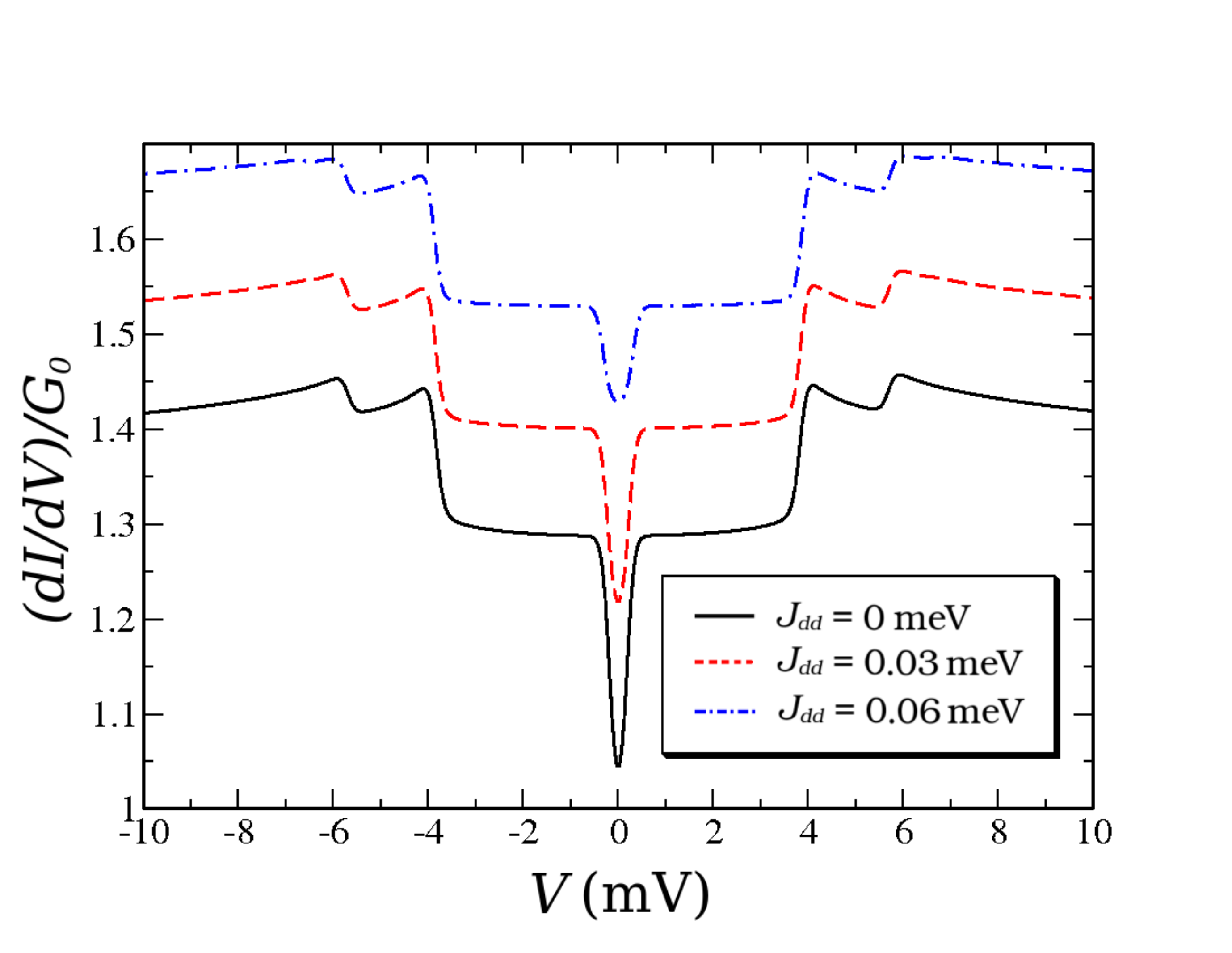}}
\caption{\footnotesize{Normalised conductance spectrum of Fe when it is exchange coupled to Co on CuN as $J_\mathrm{dd}$ is increased. 
Note that the intensity of the conductance step at 0.18~mV decreases with increased exchange coupling.}}
\label{4}
\end{figure}

As a final test for our 3rd order self-energy we consider a Co-Fe dimer with exchange coupling between the ions, a situation already investigated 
experimentally in Ref.~\cite{Otte2}. In our formalism this translates in including an additional Heisenberg-like term to $H_\mathrm{sp}$ 
\begin{equation}
\label{eq:7}
H_\mathrm{H}=J_\mathrm{dd}(\mathbf{S}_{\text{Co}}\cdot\mathbf{S}_{\text{Fe}}),
\end{equation}
\begin{figure}[t]
\centering
\resizebox{\columnwidth}{!}{\includegraphics[width=5cm,angle=-90]{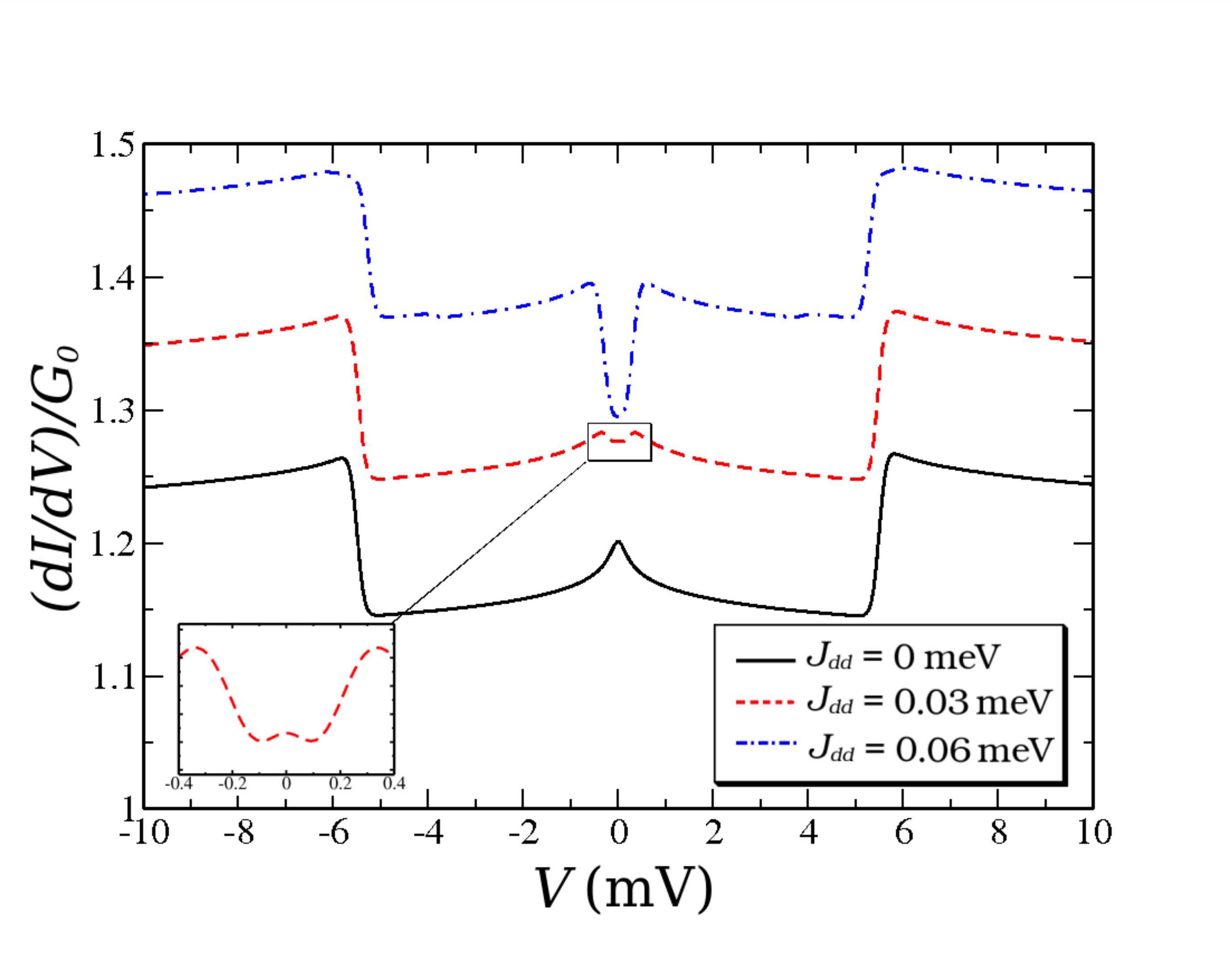}}
\caption{\footnotesize{Normalised conductance spectrum of Co when it is exchange coupled to Fe on CuN as $J_\mathrm{dd}$ is increased. 
The Co Kondo peak splits as the Fe atom acts as an effective magnetic field. The insert zooms in the zero-bias region.}}
\label{3}
\end{figure}
where $J_\mathrm{dd}$ is the exchange coupling constant. Figures \ref{3} and \ref{4} show the calculated 
conductance spectra when the tip is positioned respectively over Co or Fe.  Note that in general $J_\mathrm{dd}$ is small so that any changes in 
the electronic levels of the combined Co-Fe system are not resolved in the mV range and can be neglected \cite{Hurley}). In general we notice
a change in the conductance spectrum of each atom as they are brought closer together, i.e. as $J_\mathrm{dd}$ increases. The conductance 
step at 0.18~mV for Fe decreases with increasing $J_\mathrm{dd}$, while Fe itself acts as an effective magnetic field that splits the zero-bias Kondo 
resonance of the Co spectrum. These effects are both observed in the experiments of Ref.~\cite{Otte2}. Notably Fe does not simply act 
as a source of magnetic field on the Co, as seen in the inset of figure~\ref{3} for $J_\mathrm{dd}=0.03$~meV. In fact one can clearly observe an 
additional Kondo peak emerging at zero-bias in between the principally split peaks. This is a unique feature of the exchange coupling between 
Co and Fe. In fact there are now $(2S_{\text{Co}}+1)\times(2S_{\text{Fe}}+1)=20$ eigenvalues and additional allowed transitions appear at each 
atomic site. For instance for large $J_\mathrm{dd}$ the zero-bias region becomes completely dominated by a conductance dip. This originates from 
the opening of a spin transition between the ground state at -5.686~meV and the first excited state at -5.379~meV. Such a transition has a spectral 
intensity much larger than that of the Kondo resonance, which therefore disappears from the spectrum.

In conclusion we have studied the effects of including 3rd order contributions in the electron-spin coupling to the conductance
spectra of transition metal atoms adsorbed on a CuN substrate. In particular we have derived a close expression for the 3rd order 
electron-spin self-energy within the NEGF formalism and a single band tight-binding model incorporating 
local Heisenberg exchange to quantum spins. Two main features in the conductance spectra arise from our formalism, namely a 
logarithmic decay of the conductance as a function of bias and zero-bias Kondo resonances. We obtain a good agreement with available 
experimental data for both isolated Co and Fe atoms and for an exchanged coupled Co-Fe dimer. Importantly the low computational effort 
needed by our method makes it a valuable alternative to full many-body treatments in describing spin inelastic phenomena at the atomic level. 
We also believe that our proposed scheme is amenable to be combined with first principles methods, i.e. it can form the basis for a fully quantitative 
theory of spin scattering in nanostructures.

This work is sponsored by the Irish Research Council for Science, Engineering \& Technology (IRCSET). Computational resources have been provided by the Trinity Centre for High Performance Computing (TCHPC).

\end{document}